\documentclass[aps,prx,floatfix,twocolumn,nofootinbib,superscriptaddress]{revtex4-2}
\usepackage[dvipsnames]{xcolor}

\usepackage{graphicx,
            amsmath,
            braket,
            xspace,
            lipsum}

\begin{document}

\title{Repeated ancilla reuse for logical computation on a neutral atom quantum computer}
\author{J.~A.~Muniz}
\affiliation{Atom Computing, Inc.}

\author{D.~Crow}
\affiliation{Atom Computing, Inc.}

\author{H.~Kim}
\affiliation{Atom Computing, Inc.}

\author{J.~M.~Kindem}
\affiliation{Atom Computing, Inc.}
\author{W.~B.~Cairncross}
\affiliation{Atom Computing, Inc.}
\author{A.~Ryou}
\affiliation{Atom Computing, Inc.}

\author{T.~C.~Bohdanowicz}
\affiliation{Atom Computing, Inc.}
\author{C.-A.~Chen}
\affiliation{Atom Computing, Inc.}
\author{Y.~Ji}
\affiliation{Atom Computing, Inc.}
\author{A.~M.~W.~Jones}
\affiliation{Atom Computing, Inc.}
\author{E.~Megidish}
\affiliation{Atom Computing, Inc.}
\author{C.~Nishiguchi}
\affiliation{Atom Computing, Inc.}
\author{M.~Urbanek}
\affiliation{Atom Computing, Inc.}
\author{L.~Wadleigh}
\affiliation{Atom Computing, Inc.}
\author{T.~Wilkason}
\affiliation{Atom Computing, Inc.}

\author{D.~Aasen}
\affiliation{Microsoft Quantum}
\author{K.~Barnes}
\affiliation{Atom Computing, Inc.}
\author{J.~M.~Bello-Rivas}
\affiliation{Microsoft Quantum}
\author{I.~Bloomfield}
\affiliation{Atom Computing, Inc.}
\author{G.~Booth}
\affiliation{Atom Computing, Inc.}
\author{A.~Brown}
\affiliation{Atom Computing, Inc.}
\author{M.~O.~Brown}
\affiliation{Atom Computing, Inc.}
\author{K.~Cassella}
\affiliation{Atom Computing, Inc.}
\author{G.~Cowan}
\affiliation{Atom Computing, Inc.}
\author{J.~Epstein}
\affiliation{Atom Computing, Inc.}
\author{M.~Feldkamp}
\affiliation{Atom Computing, Inc.}
\author{C.~Griger}
\affiliation{Atom Computing, Inc.}
\author{Y.~Hassan}
\affiliation{Atom Computing, Inc.}
\author{A.~Heinz}
\affiliation{Atom Computing, Inc.}
\author{E.~Halperin}
\affiliation{Atom Computing, Inc.}
\author{T.~Hofler}
\affiliation{Atom Computing, Inc.}
\author{F.~Hummel}
\affiliation{Atom Computing, Inc.}
\author{M.~Jaffe}
\affiliation{Atom Computing, Inc.}
\author{E.~Kapit}
\affiliation{Atom Computing, Inc.}
\affiliation{Department of Physics, Colorado School of Mines, Golden, CO}
\author{K.~Kotru}
\affiliation{Atom Computing, Inc.}
\author{J.~Lauigan}
\affiliation{Atom Computing, Inc.}
\author{J.~Marjanovic}
\affiliation{Atom Computing, Inc.}
\author{M.~Meredith}
\affiliation{Atom Computing, Inc.}
\author{M.~McDonald}
\affiliation{Atom Computing, Inc.}
\author{R.~Morshead}
\affiliation{Atom Computing, Inc.}
\author{S.~Narayanaswami}
\affiliation{Atom Computing, Inc.}
\author{K.~A.~Pawlak}
\affiliation{Atom Computing, Inc.}
\author{K.~L.~Pudenz}
\affiliation{Atom Computing, Inc.}
\author{D.~Rodr\'iguez~P\'erez}
\affiliation{Atom Computing, Inc.}
\author{P.~Sabharwal}
\affiliation{Atom Computing, Inc.}
\author{J.~Simon}
\affiliation{Department of Physics and Department of Applied Physics, Stanford University, Stanford, California}
\author{A.~Smull}
\affiliation{Atom Computing, Inc.}
\author{M.~Sorensen}
\affiliation{Atom Computing, Inc.}
\author{D.~T.~Stack}
\affiliation{Atom Computing, Inc.}
\author{M.~Stone}
\affiliation{Atom Computing, Inc.}
\author{L.~Taneja}
\affiliation{Atom Computing, Inc.}
\author{R.~J.~M.~van de Veerdonk}
\affiliation{Atom Computing, Inc.}
\author{Z.~Vendeiro}
\affiliation{Atom Computing, Inc.}
\author{R.~T.~Weverka}
\affiliation{Atom Computing, Inc.}
\author{K.~White}
\affiliation{Atom Computing, Inc.}
\author{T.-Y.~Wu}
\affiliation{Atom Computing, Inc.}
\author{X.~Xie}
\affiliation{Atom Computing, Inc.}
\author{E.~Zalys-Geller}
\affiliation{Atom Computing, Inc.}
\author{X.~Zhang}
\affiliation{Atom Computing, Inc.}

\author{J.~King}
\affiliation{Atom Computing, Inc.}
\author{B.~J.~Bloom}
\affiliation{Atom Computing, Inc.}
\author{M.~A.~Norcia}
\affiliation{Atom Computing, Inc.}
\email{matt@atom-computing.com}

\begin{abstract}
\noindent 
Quantum processors based on neutral atoms trapped in arrays of optical tweezers have appealing properties, including relatively easy qubit number scaling and the ability to engineer arbitrary gate connectivity with atom movement.  However, these platforms are inherently prone to atom loss, and the ability to replace lost atoms during a quantum computation is an important but previously elusive capability.  Here, we demonstrate the ability to measure and re-initialize, and if necessary replace, a subset of atoms while maintaining coherence in other atoms. This allows us to perform logical circuits that include single and two-qubit gates as well as repeated midcircuit measurement while compensating for atom loss.  We highlight this capability by performing up to 41 rounds of syndrome extraction in a repetition code, and combine midcircuit measurement and atom replacement with real-time conditional branching to demonstrate heralded state preparation of a logically encoded Bell state. Finally, we demonstrate the ability to replenish atoms in a tweezer array from an atomic beam while maintaining coherence of existing atoms -- a key step towards execution of logical computations that last longer than the lifetime of an atom in the system.  

\end{abstract}
\maketitle
\section{Introduction}\label{introduction}

Many prominent protocols for error-corrected quantum computing rely on the ability to measure the state of a subset of qubits -- ``ancilla" qubits -- while maintaining quantum coherence in other ``data" qubits \cite{shor1995scheme, knill1997theory}.  This capability is referred to as midcircuit measurement (MCM).  Motivated by this need, several demonstrations of MCM have recently been performed in systems comprised of individually-controlled neutral atoms trapped within optical tweezers \cite{graham2023mid, Deist2022mid, singh2023mid, norcia2023midcircuit, lis2023mid, bluvstein2023logical, finkelstein2024universal}.  In order to perform fault tolerant circuits with multiple rounds of MCM, it is advantageous to be able to reuse ancilla qubits after reading out their state, requiring MCM to be both nondestructive and accompanied by reset of the qubit state and motional state of the measured atoms.  State readout of neutral atoms -- including some previous demonstrations of MCM \cite{singh2023mid, bluvstein2023logical} -- is often destructive, relying on the removal of one qubit state to obtain state-selectivity.  Other MCM protocols may be fundamentally compatible with ancilla reuse \cite{Deist2022mid, lis2023mid, finkelstein2024universal}, but ancilla reuse within a quantum circuit involving a universal set of gate capabilities has not previously been demonstrated.  This work demonstrates quantum circuits that contain repeated MCM with reset and reuse of ancilla qubits.  

\begin{figure*}
    \centering
        \includegraphics[width=2.0\columnwidth]{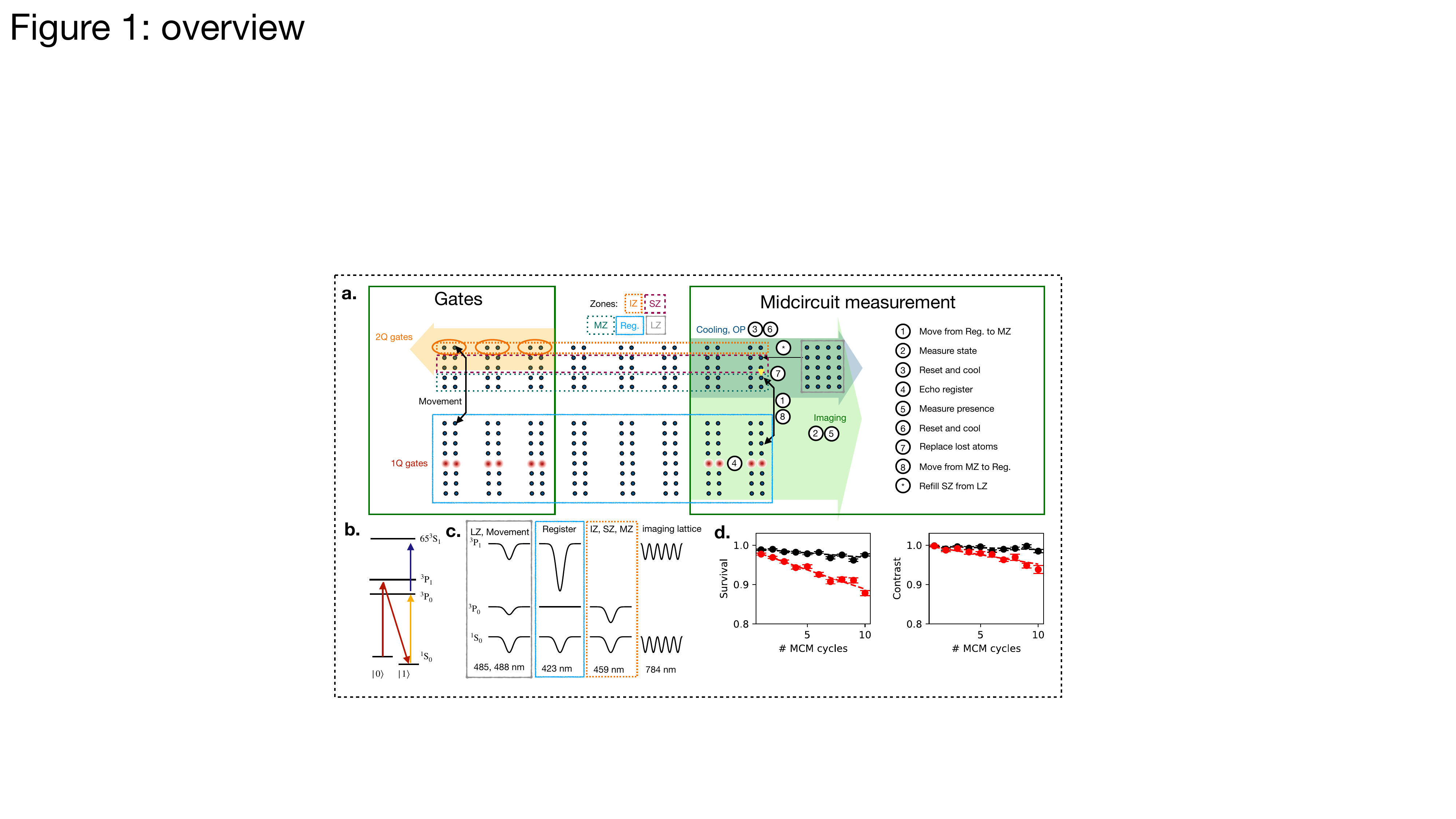}
    \caption{\textbf{a.} Zoned architecture for universal quantum computing.  Optical tweezers are divided into sub-arrays, or zones, labeled interaction zone (IZ), storage zone (SZ), measurement zone (MZ), register (Reg.), and loading zone (LZ). Atoms can be handed from tweezers to a collocated cavity-enhanced lattice (not shown) for readout. Two-qubit (2Q) controlled-Z (CZ) gates are performed in parallel on all populated pairs of tweezers within the IZ (2Q gate lasers are represented by orange arrow). Site-resolved arbitrary single-qubit (1Q) gates are performed within the register (1Q gate lasers are represented by red circles). Arbitrary connectivity for 2Q gates is provided by moving selected atoms to desired positions within the IZ. Midcircuit measurement (MCM) allows state-selective, nondestructive imaging of the atoms within the IZ, SZ, and MZ, while maintaining coherence within the register. MCM is followed by cooling, qubit reset, and replacement of missing atoms with atoms from the SZ. The imaging laser is represented by the green arrow, while the lasers used for optical pumping and cooling are represented by the blue arrow. A full MCM cycle is depicted, and described in detail in the main text.  
    \textbf{b.} Selected energy levels of $^{171}$Yb. Qubit states are defined by the nuclear spin of the ground state $^1$S$_0$. Single-qubit gates are implemented using two-photon Raman transitions near-detuned from the $^3$P$_1$ manifold. Two-qubit gates are performed by sequential excitation from $\ket{1}$ to a high-lying Rydberg state, through a metastable $^3$P$_0$ state \cite{muniz2024gates}.   
    \textbf{c.} Qualitative polarizabilities of relevant states at wavelengths used for generating the different arrays, represented schematically as depths of trapping potentials.  Within the LZ and movement tweezers, as well as the imaging lattices, $^1$S$_0$ and $^3$P$_1$ have equal polarizability. Within the register, $^3$P$_1$ has large polarizability, while $^3$P$_0$ is un-trapped. In the tweezers that form the IZ, SZ and MZ, $^1$S$_0$ and $^3$P$_0$ have equal polarizability. 
    \textbf{d.} Performance of MCM. Contrast and atom loss in the register are extracted from a Ramsey sequence containing a variable number of MCM cycles. Red markers correspond to the full MCM sequence. Black markers correspond to the same timing and operations (including lattice handoffs), but without imaging, cooling, or OP light applied.  In total, the MCM cycle contributes a fractional loss of 0.0106(7) and a contrast loss of 0.0049(7) among surviving atoms. 
    }
    \label{fig:mcm}
\end{figure*}

Even MCM protocols that are nominally nondestructive will suffer from occasional loss of ancilla qubits, either from the MCM process itself, or from idle or operational losses in other parts of the circuit.  To compensate for this loss, it is advantageous to be able to replace missing atoms after MCM. Multiple rounds of MCM can also be performed by replacing lost ancilla atoms in destructive MCM protocols, \cite{finkelstein2024universal}, though the much higher required rate of replacement presents challenges for large numbers of atoms or rounds of MCM. Here, we replace ancilla atoms that are occasionally lost with atoms from a reservoir region. Because reservoir-based approaches will eventually run out of atoms, very long circuits also require a way to reload the reservoir while maintaining coherence in data qubits. We demonstrate this capability by refilling the reservoir from a spatially separated magneto-optical trap (which itself is reloaded from an atomic beam), while maintaining coherence in qubits already within the system.  

We showcase the ability to perform repeated reuse of measured atoms by performing repeated cycles of a repetition code, where ancilla atoms are reused between rounds of execution.  This code is configured to swap the role of data and ancilla qubits on each cycle, allowing us to perform a delayed erasure conversion and correct for occasional loss of ancilla atoms through replacement from a reservoir while preserving logical states \cite{2024AtomQEC,chow2024leakage,perrin2025quantumerrorcorrectionresilient}.  We also perform up to 41 cycles of syndrome extraction while replacing ancillae as they are lost, observing error detection rates that are roughly constant versus the repetition index.

By combining MCM with ancilla reuse and classical branching, we demonstrate the ability to perform repeated attempts at fault-tolerant state preparation of logical Bell states, where the repetitions are performed until measurements of parity-check qubits indicate successful preparation of the desired state. Such capabilities form the basis of scalable state preparation, as multiple blocks can be built up one at a time, and stored as further blocks are prepared.  

\section{Zoned architecture and midcircuit measurement}\label{mcm_description}

In this work, we demonstrate a zone-based quantum processor \cite{bluvstein2023logical, 2024AtomQEC} that is capable of performing universal quantum computation as well as midcircuit measurement with qubit reset and reloading. This is facilitated by a combination of trapping wavelengths used in spatially separated zones to enable the local application of key operations.  Our system uses nuclear-spin qubits formed within the $^1$S$_0$ ground states ($\ket{0} \equiv \ket{F=1/2, m_F = -1/2}$ and $\ket{1} \equiv \ket{F=1/2, m_F = +1/2}$) of $^{171}\text{Yb}$ atoms \cite{2024AtomQEC}. Quantum gates and atom movement are performed using optical tweezer arrays, while high-fidelity, non-destructive state-selective readout is performed by transferring atoms to collocated cavity-enhanced optical lattices \cite{norcia2024iterative}.  Imaging is performed by collecting photons scattered from the relatively narrow (180~kHz linewidth) $\ket{1}$ to $^3$P$_1$ $F=3/2, m_F = 3/2$ transition at 556~nm on a sensitive camera. Importantly, our imaging technique \cite{norcia2023midcircuit, huie2023repetitive} enables unambiguous determination of both the qubit state and atom presence. Further, key gate errors are converted to atom loss \cite{2024AtomQEC}. The combination of these features allows us to perform (delayed) erasure conversion \cite{Wu2022erasure, chow2024leakage,perrin2025quantumerrorcorrectionresilient}.  

The zones and their functions are illustrated in Fig.~\ref{fig:mcm}(a), and key atomic states and gate lasers are shown in Fig.~\ref{fig:mcm}(b). The \textit{register} is a region of 128 optical tweezers arranged in eight pairs of adjacent columns (16 sites per row), separated by gaps that facilitate atom movement. The register holds atoms during a quantum circuit, and arbitrary site-selective single-qubit (1Q) gates are applied in the register with a pair of Raman beams \cite{barnes2022assembly,bluvstein2023logical,muniz2024gates,2024AtomQEC}. The 423~nm wavelength of the register tweezers induces substantial light-shifts on the imaging transition, which suppresses scattering when atoms in other zones are imaged (see Appendices~\ref{mcm_sequence} and \ref{3_level_system} for details). This is possible because both register tweezers and cavity lattices are present during mid-circuit measurements.

A second group of 80 tweezers operate at 459~nm and have a matching column distribution to the register. These tweezers serve multiple functions: a single row forms the \textit{interaction zone} (IZ), trapping atoms during our two-qubit (2Q) controlled-Z (CZ) entangling gates via sequential excitation to a Rydberg state \cite{muniz2024gates}; two rows make up the \textit{measurement zone} (MZ), where atoms are placed to be measured during MCM; and the final two rows function as the \textit{storage zone} (SZ), which acts as a nearby reservoir to replenish any missing atoms in the MZ after MCM. The 459~nm tweezers provide equal light-shifts to the $^1$S$_0$ ground and $^3$P$_0$ states \cite{hohn2023state}, which is key for executing the 2Q gates \cite{muniz2024gates}.  

Finally, the \textit{loading zone} (LZ) consists of 75 sites at 483~nm \cite{norcia2023midcircuit}. This densely packed array's main purpose is to load and prepare fresh single atoms that are transferred from a magneto-optical trap situated 300~mm below, as detailed in \cite{norcia2024iterative}.

Arbitrary gate connectivity across the system is enabled by atom movement \cite{bluvstein2023logical, 2024AtomQEC}. These movements are executed using a set of 488~nm tweezers, controlled by a pair of crossed acousto-optical deflectors (AODs), utilizing the empty ``highways" situated between column pairs within these arrays to move atoms while minimizing disturbances on qubits in the register. All atom movements occur while the cavity lattice is off.

MCM in our zoned architecture relies on the ability to image the atoms in the MZ and SZ with high fidelity, perform cooling and state preparation for atoms in the MZ and SZ after imaging, and move atoms from the SZ to the MZ to correct for detected loss, while minimizing loss and retaining the coherence of superposition states within atoms in the register. Imaging is performed using a global beam that illuminates all atoms (due to geometrical constraints). Cooling after imaging, as well as optical pumping, are implemented using local beams that illuminate only the IZ, SZ and MZ. The large ($\simeq 100$~MHz) shifts induced by the register tweezers dramatically reduce scattering among register atoms from either the global imaging beam and stray light from the local beams.

A typical \textit{MCM cycle} is described in Fig.~\ref{fig:mcm}(a). It begins by moving subsets of ancilla qubits from the register to the MZ tweezers. Atoms from the MZ and SZ are then transferred from the 459~nm tweezers to the lattice for imaging. A first image serves to identify atoms in $\ket{1}$. This is immediately followed by a \textit{MCM reset} cycle consisting of gray-molasses cooling \cite{jenkins2022yb} and optical pumping that prepare atoms in the MZ and SZ into state $\ket{1}$. A spin echo is performed on atoms in the register to cancel shifts from the MCM process \cite{googleAI_repcode_2021,norcia2023midcircuit} (see Appendices~\ref{mcm_sequence} and \ref{3_level_system} for details). A second image of state $\ket{1}$ confirms the presence of atoms regardless of their initial internal state, followed by a second reset to cool and prepare MZ and SZ atoms into state $\ket{0}$, which is not excited in our 2Q gates \cite{muniz2024gates}. 

In Fig~\ref{fig:mcm}(d) we show the typical atom survival and maintenance of coherence for atoms in the register during MCM by inserting repeated MCM cycles in a Ramsey sequence. We observe atom loss of 0.0106(7) per cycle, and an site-averaged contrast loss of 0.0049(7) for atoms that survive in the register. Separately, we measure the survival of atoms in the MZ under MCM, and find a per-cycle loss of 0.005(2). We discuss the origin of this loss in Appendix \ref{3_level_system}.


  
After the MCM block, the second MCM image is analyzed in real-time by a software service that determines atom presence in the MZ and SZ, and programs the movement tweezers to fill detected vacancies within the MZ with atoms from the SZ. We call such movements from filled sites to empty sites \textit{conditional movements}. Further details can be found in Appendix~\ref{mcm_sequence}.

\begin{figure}
    \includegraphics[width=1.0\columnwidth]{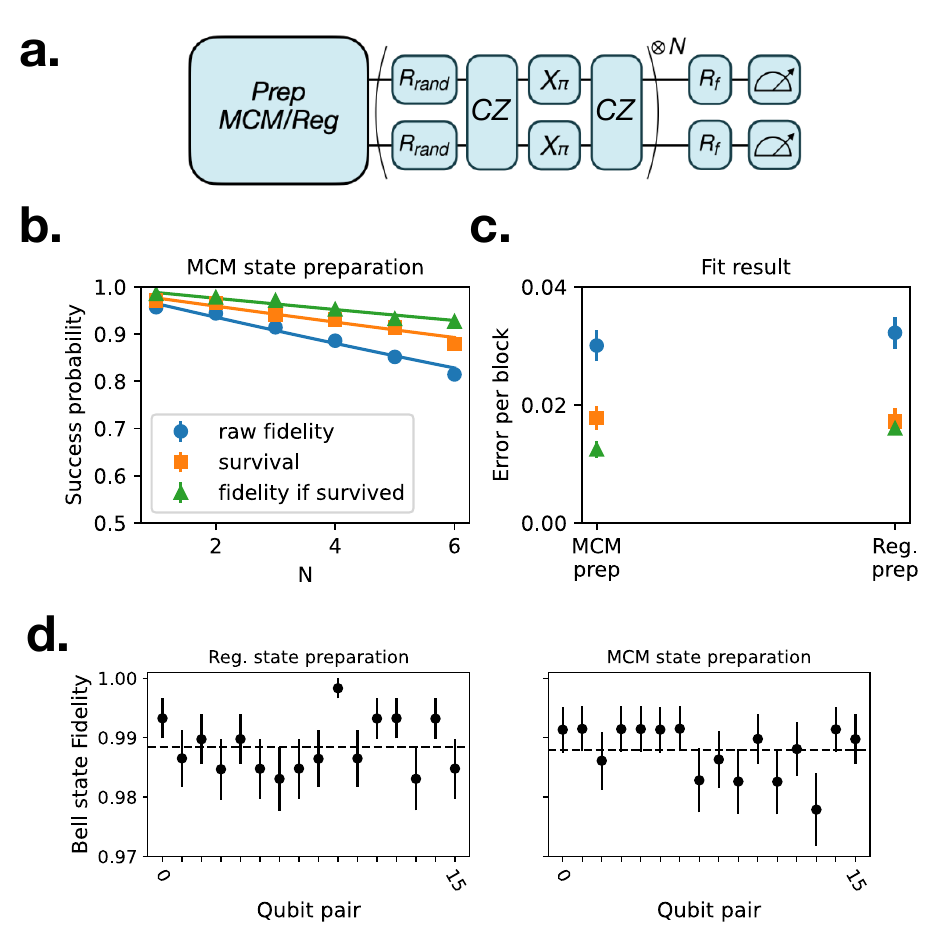}
    \caption{Gate performance following MCM reinitialization. 
    \textbf{a.} The Global-Echo Randomized Benchmarking (GERB) sequence consists of $N$ repeated blocks of 2Q and random 1Q gates performed on pairs of qubits, followed by a pre-computed operation to bring population back to the $\ket{11}$ state in the absence of errors.  We perform this sequence using atoms prepared using an MCM cycle in the MZ (MCM prep), or with atoms prepared directly in the register (Reg. prep).  
    \textbf{b.} Success probability versus $N$ for MCM prep. By measuring both qubit states and survival \cite{2024AtomQEC, muniz2024gates}, we separately extract total success probability (blue), cases where both atoms survive (orange), and the probability of measuring the correct two-qubit state when both atoms survive (green). \textbf{c.}  Error rate per block, as extracted from exponential fits in subfigure b (and similar data for register prep), indicating compatible performance between the two preparation methods is shown on the right.  
    \textbf{d.} Bell state fidelities for atoms initialized using register preparation (left) and MCM state preparation (right), plotted versus qubit pair indices for qubits within the top two rows of the register.
    }
    \label{fig:ancilla_reuse}
\end{figure}

\section{Atom reuse following MCM}\label{anc_reuse}
In order to reuse atoms within a circuit following an MCM cycle, they must be reinitialized into a specific internal state (we use $\ket{0}$) and be sufficiently cold to enable high-fidelity gates. The imaging itself leaves the atoms slightly too warm for this, so we apply gray-molasses cooling after imaging to reduce the radial motional quantum numbers to $n\simeq 0.5$, which is compatible with high-fidelity gates in our system \cite{2024AtomQEC, muniz2024gates}. Here, we perform two gate performance benchmarks that compare atoms prepared in our standard register imaging and state preparation protocol ``register preparation" (non-MCM as in Ref.~\cite{2024AtomQEC}) versus cases where the atoms are prepared using an MCM cycle ``MCM preparation". 

In Fig~\ref{fig:ancilla_reuse}(a), we perform a measurement that is sensitive to both 1Q and 2Q gate fidelities using a variation of the Global-Echo Randomized Benchmarking (GERB) protocol. In this protocol, concatenated circuit blocks each contain a random 1Q operation and two CZ gates separated by an echo pulse, as depicted in Fig.~\ref{fig:ancilla_reuse}(a) and detailed in \cite{muniz2024gates,2024AtomQEC}. This method is similar to those used to characterize 2Q gate performance in other cold-atom systems \cite{evered_high-fidelity_2023}. A final 1Q operation brings pairs of atoms to the $\ket{11}$ state in the absence of errors.  
By observing the decay of population returned to $\ket{11}$ (and also of atom-pair survival) versus the number of blocks, we extract the combined error rate for the operations in the block (Fig.~\ref{fig:ancilla_reuse}(b)). We compare these error rates for register preparation and MCM preparation in Fig.~\ref{fig:ancilla_reuse}(b), and find consistent performance between the two cases.  

Our second benchmark characterizes our ability to generate the Bell state $\ket{\Phi_+} = (\ket{00}+\ket{11})/\sqrt{2}$ after each of the two preparation methods using modified versions of the circuits in \cite{cruz2019efficient}.  In the MCM case, we prepare the atoms in the MZ, then move them to the register 
and then apply the same Bell state generation circuit as when the atoms are prepared directly in the register.  
The fidelities (as defined in \cite{moses_2023_ionQPU}) for both cases are shown in Fig.~\ref{fig:ancilla_reuse}(c). We observe an average Bell state fidelity of 98.8(1)\% for both preparations. These Bell state fidelities are post-selected on atom survival \cite{2024AtomQEC}, and so do not include additional loss associated with MCM cycle itself, which is characterized above. Additional loss in CZ gates for the MCM-prepared case is bounded by the GERB benchmark data of Fig.~\ref{fig:ancilla_reuse}(b).  

\section{Repetition code with atom reuse and replacement}\label{rep_code}

\begin{figure*}
    \centering
        \includegraphics[width=2.0\columnwidth]{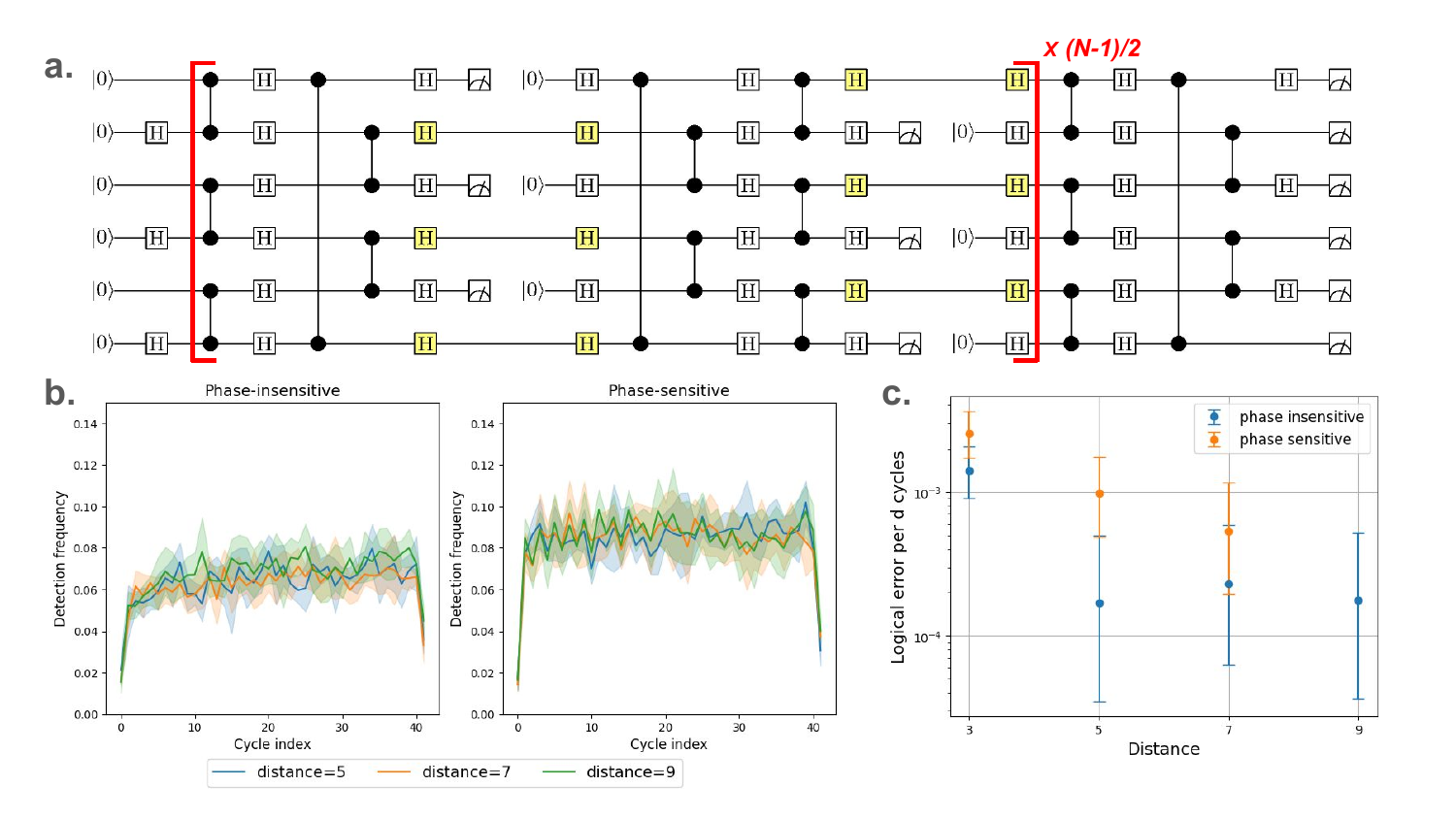}
    \caption{\textbf{a.} Distance 3, $N$ cycle, phase-sensitive repetition code in the $H$ and CZ basis. Information is encoded in the physical state of $N$ qubits which undergo cyclic parity checks with $N$ ancilla qubits. Physically, this circuit is compiled in the machine native gate set: ($R_z(\phi)$, $S_x$, CZ) and an additional $X$ gate is applied to the non measured qubits to echo noise during MCM. The highlighted $H$ gates are removed for the phase-insensitive case. The portion of the circuit within the red brackets is repeated $(N-1)/2$ times.  \textbf{b.} Experimental results for the average detector frequency for 41 cycles and different code distance with and without phase sensitivity. In each case the detectors are calculated from the measurements for each circuit and average (solid line) and standard deviation (shaded area) of ~600 shots is calculated. For the phase-sensitive case we can see slightly elevated detection, as expected from the additional 1Q gates and sensitivity to phase errors occurring during the MCM and reinitialization steps.    
    \textbf{c.} Demonstration of error suppression as a function of code distance computed for experiments with number of cycles equal to distance. Each datapoint consists of at least 11000 shots. The errorbars show 95\% confidence intervals. Data used to generate this plot can be found in Appendix \ref{rep_code_extra} Table \ref{tab:rep_data}.
    }
    \label{fig:rep_code}
\end{figure*}
We demonstrate the integration of MCM and atom reuse with a circuit-based error correction protocol. We encode a single classical logical bit in a repetition code using $N$ physical qubits. We call this the distance of the code. We use $N$ additional physical qubits to simultaneously perform syndrome extraction and loss detection through swap-based syndrome extraction. This circuit is constructed by adding SWAP gates to a standard syndrome extraction circuit and then compiling it into our native gateset \cite{2024AtomQEC,chow2024leakage}. The final circuit has the same number of 2Q gates as a standard syndrome extraction circuit. Although the standard repetition code requires only $N-1$ parity checks, we implement a ``ring code" or 1D toric code by adding one additional check. This additional check facilitates constructing a circuit in which a physical qubit is active for at most two cycles before being measured, reinitialized, and if necessary, replaced. For further information and raw data used to generate plots see Appendix~\ref{rep_code_extra}.

We focus here on a bit-flip variant of the repetition code, though in our native gateset a phase-flip variant differs only by the insertion of 1Q gates that resolve to identity. Typically, this type of repetition code places each qubit in a computational basis state during MCM and so is insensitive to phase errors induced by MCM on data qubits. To make the code sensitive to such phase errors (which must be small for a quantum error correcting code), we also characterize a second variant of the code that has a pair of Hadamard gates added to data qubits surrounding each MCM, placing the data qubits in a superposition state during MCM. We refer to these two variants of the code as \emph{phase-insensitive} and \emph{phase-sensitive}, respectively (Fig. \ref{fig:rep_code}(a)).


We performed repetition code memory experiments for various distances and numbers of cycles. In particular, we scanned odd distances between 3 and 9 while keeping the number of rounds equal to the distance. For these experiments, the results were decoded using minimum-weight perfect matching implemented using PyMatching \cite{Higgott2025sparseblossom}. The matching graph was modified on a per-shot basis to account for atom loss. For more information, see Appendix~\ref{rep_code_extra}. A plot of logical failure rate versus distance is shown in Fig.~\ref{fig:rep_code}(c). Although we see improvement of logical error rate with increasing distance, we also see evidence of a plateau, particularly for the phase-insensitive data. In this work, we prioritize the demonstration of device functionality and lack the data to seriously investigate this plateau. We leave a more thorough investigation of logical performance to future work.

In addition to tracking logical performance, we conducted separate memory experiments for odd distances between 5 and 9 while repeating the syndrome extraction procedure for 41 rounds. For these experiments, we compute the frequencies of detection events \cite{fowler2014scalable,googleAI_repcode_2021} in order to examine the system behavior. These `detectors' represent the overall parities between consecutive syndrome checks on corresponding pairs of data qubits. Unlike bare syndrome measurements, detectors localize regions of error sensitivity.
In Fig.~\ref{fig:rep_code}(b), we plot the average detection frequency per cycle for the phase-insensitive and phase-sensitive versions of the repetition code, respectively. We see the detection frequencies in the bulk remaining relatively flat across multiple rounds. In both cases we can see that the detector probability is independent of the distance (within the error bars). 

\section{Heralded fault-tolerant state preparation with conditional repetition}\label{repeat_until}

\begin{figure}
    \includegraphics[width=1.0\columnwidth]{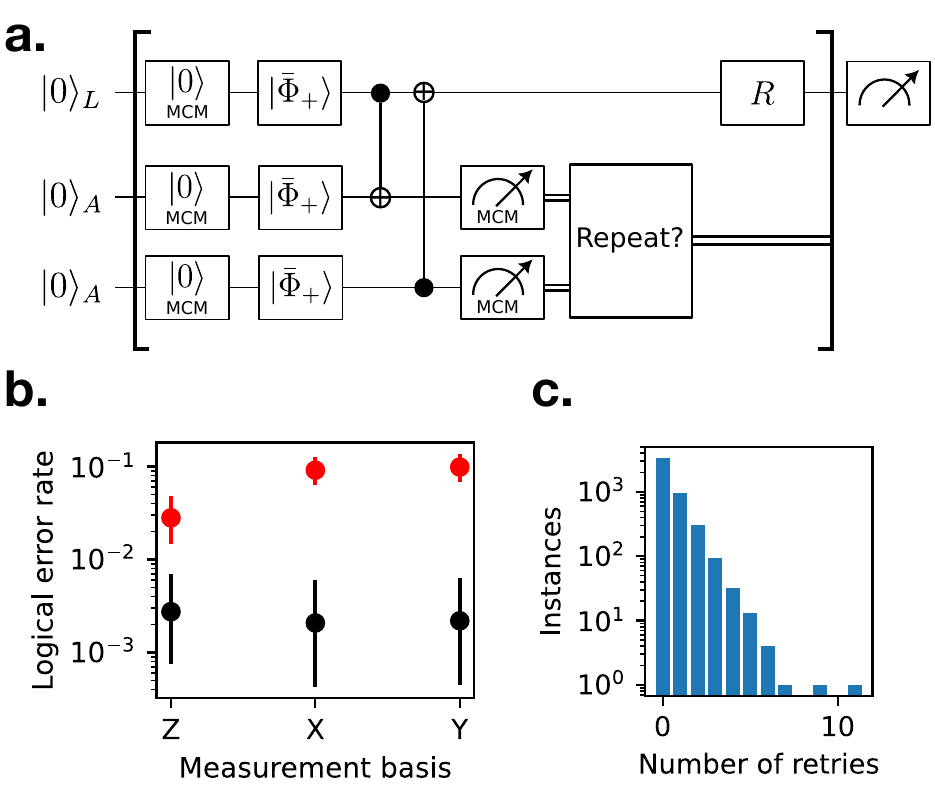}
    \caption{\textbf{a.} Fault-tolerant state preparation of logical Bell states encoded in the [[4,2,2]] code through conditional retries. At the beginning of each state preparation attempt, data and ancilla qubits are reset via an MCM operation ($\ket{0}_{\text{MCM}}$) and a set of logical Bell states $\ket{\bar{\Phi}_+}=\ket{\Phi_+}\otimes\ket{\Phi_+}$ are prepared in the data and ancilla blocks (four physical qubits each). After preparation ancillas are entangled with the data qubits and then measured using MCM.  After a decoder assesses if the correct parity was measured, the data qubits are rotated to the desired basis using the rotation $R$. If the parity was incorrect, another attempt is performed. If the parity was correct, the retry loop exits and the logical state is measured.  
    \textbf{b.} Logical error rate on different measurement basis for instances where the decoder succeeded using the final measurement (black), and when it did not using the reset measurement $\ket{0}_{\text{MCM}}$ (red). Error bars represent the 95\% confidence interval. Data used to create this plot can be found in Appendix \ref{422_prep}, Table \ref{tab:belldata}.
    \textbf{c.} Histogram representing the number of retries required to obtain a successful parity check. The average number is 0.44.  
    }
    \label{fig:repeat_until}
\end{figure}

Fault-tolerant (FT) preparation of high-quality resource states is an essential feature of practical quantum computation. Leveraging our MCM and ancilla reuse scheme, we demonstrate heralded preparation of logical Bell states. This is achieved by iteratively repeating a [[2,1,2]] distillation protocol encoded into [[4,2,2]] codes. We prepare logical Bell pairs $\ket{\bar{\Phi}_+}$ on a data block and two ancillary blocks --- one to identify bit-flip errors and another to identify phase errors. We repeat the distillation protocol until a set of MCM measurements on the ancillary blocks indicates successful state preparation of a target logical state on the data block. The defining characteristic of this demonstration is the execution of a conditional loop that repeats until the success criterion is met. By using MCM to measure the ancilla blocks, we preserve the logical state on the data block, which remains available for use after measurement and classical logic are applied, as illustrated in Fig.~\ref{fig:repeat_until}(a). Details about the [[4,2,2]] code and the distillation protocol as well as the data used to create Fig. \ref{fig:repeat_until}(b) can be found in Appendix~\ref{422_prep}.

After the ancilla parity check succeeds and the state preparation loop exits, we observe a basis-averaged encoded failure rate of $0.4^{(4)}_{(2)}$\%, indicating trials where the prepared state decodes to the incorrect logical state. Failure rates in the three measurement bases are shown as black markers in Fig.~\ref{fig:repeat_until}(b). Additionally, in $0.3\%$ of trials, the decoder detects an error that cannot be corrected in post-processing. We observe that on average 1.44 attempts of FT state preparation and distillation are required to exit the loop, as shown in the histogram in Fig.~\ref{fig:repeat_until}(c). 

The reset operation performed at the beginning of each state preparation attempt is implemented as an MCM cycle that contains an image of all qubits, so we can compare the logical failure rate of attempts where the ancilla did not have the correct parity (and the loop did not exit) to those where the correct parity was detected and the loop did exit. The red markers in Fig.~\ref{fig:repeat_until}(b) show the logical error rate for attempts where the incorrect parity was measured, with a basis-averaged logical failure rate of $0.11^{(3)}_{(2)}$, and where in $7\%$ of trials an uncorrectable error was detected. 

We can compare the fidelities from the heralded distillation of logically encoded states to the fidelities measured from distillation using unencoded Bell states.  We observe an unencoded fidelity of $97.7(3)$\%, compared to an encoded fidelity of $99.6^{(2)}_{(3)}$\% showing superior performance for the encoded case.

\section{Coherence-preservation during atom replenishment}\label{reloading_main}

\begin{figure*}
    \centering
        \includegraphics[width=2.0\columnwidth]{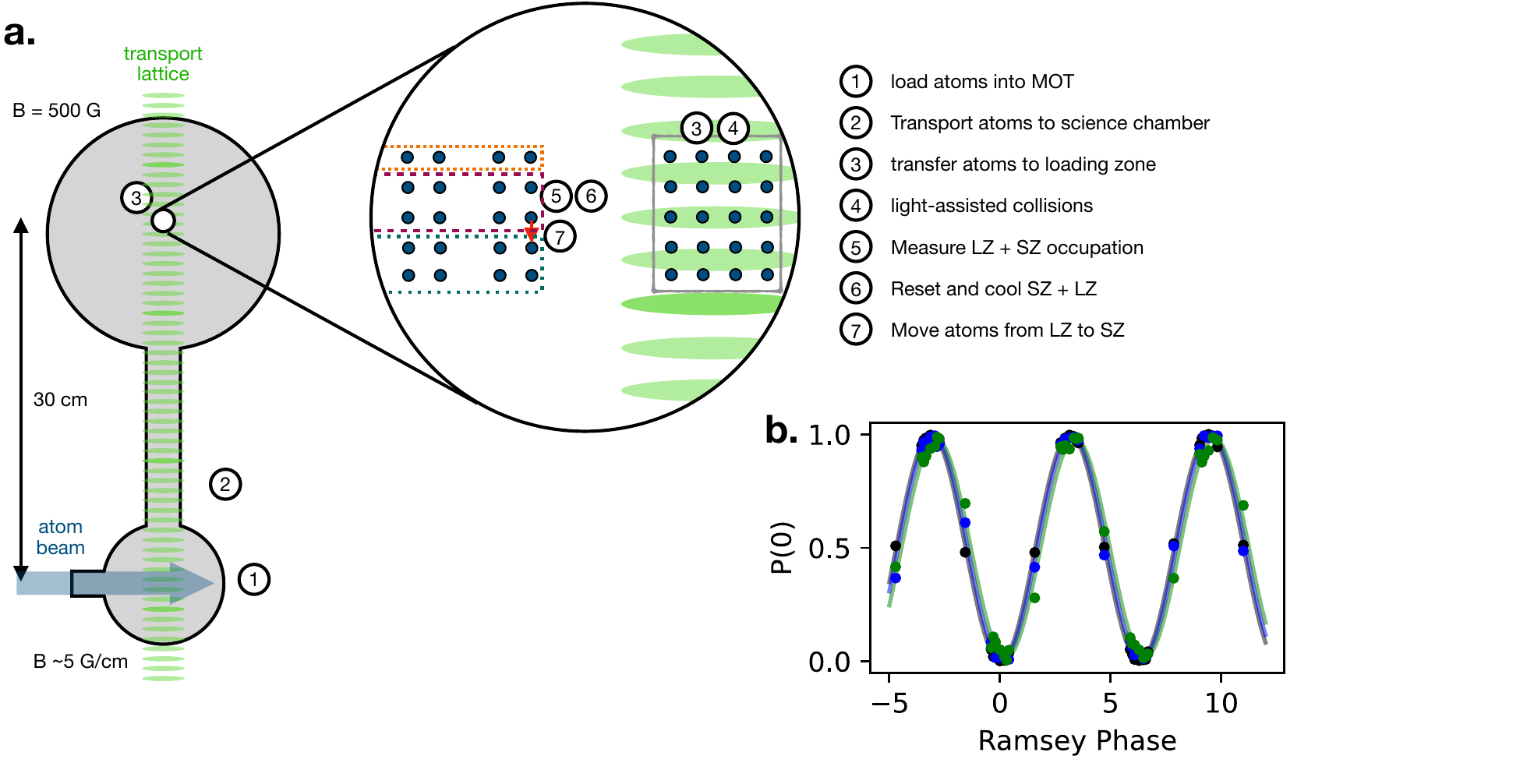}
    \caption{Replenishing atoms in tweezer arrays.  \textbf{a.} Atoms are replenished into the storage zone from a magneto-optical trap (MOT), which is in turn filled from an atomic beam.  The MOT is formed 30~cm away from the tweezer array, and then transported to the tweezer array using a moving optical lattice.  After atoms are transferred from the lattice to LZ tweezers, light-assisted collisions (LAC) are induced to project tweezer populations to zero or 1 atom.  Occupation of the LZ and SZ are determined through an MCM block, after which atoms are moved from the LZ to the SZ to fill vacancies.   All magnetic fields are static during all protocols reported here.  \textbf{b.} Coherence of atoms in the register is maintained during the atom replenishment process, as demonstrated by inserting the replenishment sequence inside a Ramsey sequence.  Ramsey interference fringes, averaged over the full array are shown for the case with the full replenishment sequence present (green), for an idle sequence of equal duration to the replenishment sequence (blue), and for a zero-hold-time sequence (black).  }
    \label{fig:reloading}
\end{figure*}

While we can refill lost atoms in the MZ with new atoms from the SZ, very long circuits will eventually deplete the population of the SZ, requiring replenishment to increase the logical lifetime beyond the lifetime of atoms in our system. Here, we refill the SZ with new atoms while maintaining coherence for atoms in the register. This capability is enabled by a combination of MCM, as demonstrated above, as well as midcircuit conditional rearrangement and the ability to form a MOT is a spatially separated region from the tweezer array, as previously demonstrated in Ref.~\cite{norcia2024iterative}.  Additionally, midcircuit replenishment requires the ability to transport atoms from the MOT to tweezer regions and perform light-assisted collisions (LAC) that provide a sub-Poissonian distribution of 0 or 1 atom in each trap \cite{schlosser2001sub}. All operations must be performed while maintaining coherence of atoms in the register. The sequence used for replenishment is shown in Fig.~\ref{fig:reloading}(a), and described in detail in Appendix~\ref{reloading}.

We first load a core-shell type MOT \cite{Lee2015coreshell} with atoms from an atomic beam source. Atoms are transferred from the MOT into an optical lattice whose sites and focus translate to move the atoms near the tweezer arrays.  Atoms are then loaded into the tweezers of the LZ, where LAC is performed. A block of imaging, cooling and optical pumping using our standard MCM protocol allows the software service to identify populated sites within the LZ, as well as empty sites within the SZ. Atoms are then conditionally moved into the SZ sites, completing the replenishment sequence. The MOT may be loaded concurrently with other operations with negligible decoherence (See Appendix~\ref{reloading}). All other steps take approximately 300~ms to complete (the exact timing depends on the number of sites to be filled and the specific paths required for rearrangement).

We characterize the degree to which coherence is maintained in the register by performing a Ramsey sequence on atoms in the register, with a replenishing sequence inserted in the decoherence-sensitive portion of the sequence. With the full replenishment sequence inserted, we observe a fitted Ramsey fringe amplitude (contrast) of 95.6(14)\% when fitting to an average over all sites in the register array, normalized to the zero-time observed contrast of 99.2(4)\% as shown in Fig.~\ref{fig:reloading}(b). The average contrast extracted for individual sites across the array is 98.1(7)\%, normalized to unity contrast, suggesting that residual phase shifts are the dominant cause of contrast loss, and may potentially be mitigated with improved calibration or echo placement. An idle sequence of duration equal to a typical replenishment sequence shows a normalized contrast of 98.0(12)\%.  

For our typical operating conditions, the fraction of SZ sites that are filled by the replenishment protocol exceeds 90\%. Because the presence of atoms in the SZ is always measured before using those atoms to refill the MZ, the reloading fraction only impacts the frequency with which replenishment must be performed, but does not contribute directly to state preparation errors in a circuit.  


\section{Discussion/conclusions}\label{conclusions}
In this work, we have demonstrated the ability to reinitialize and reuse ancilla atoms following a midcircuit measurement in a neutral-atom quantum processor. By structuring error correction cycles to periodically swap the role of data and ancilla qubits, all atoms in the array can be measured after a finite number of operations, allowing for extended error correcting circuits to be run without accumulated errors due to atom heating and loss. 
We have also shown the ability to replenish atoms in the computational array from a thermal beam source mid-circuit, in principle enabling the indefinite execution of quantum circuits on a platform where qubits have an inherently limited lifetime.  

So far, we have used these capabilities to demonstrate a classical error-correction code -- the repetition code. While the repetition code itself protects classical information (qubit populations, not coherences), the modified version we have demonstrated is also sensitive to potential phase errors during measurement, and so demonstrates the key capabilities needed for repeated execution of a true quantum error correcting code (one that corrects both amplitude and phase errors) with atom reuse.  

We have also shown how our MCM can be combined with real-time decision-making to deterministically prepare a logical state heralded by the successful parity measurements of a block of ancilla qubits. In this case, the error rate on attempts with the correct ancilla parity is well below that of attempts with the incorrect ancilla parity, and below the error rate for unencoded distillation.  

Finally, we have demonstrated the ability to replenish atoms in our tweezer arrays while maintaining coherence in existing atoms.  This realizes a key step in solving a particular hurdle for neutral-atom-based quantum computation -- the finite lifetime of atoms within the computational system -- by in principle allowing the logical state of the system to persist beyond the lifetime of individual atoms.  


\newpage
\appendix 
\section{System details}\label{system}

Atoms are loaded into the register and storage zone via the methods detailed in Ref.~\cite{norcia2024iterative}. This protocol ensures a sufficient qubit count within the register for circuit execution and establishes an initial reservoir of atoms in the SZ to replenish atom loss observed during each MCM cycle. Typically, 128 register sites and 32 storage zone sites are filled. All zones, excluding the loading zone (LZ), are arranged in eight pairs of matching columns, thereby creating ``highways" for inter-zone atom transport. A static bias magnetic field of 500~G is aligned parallel to the array rows. A razor blade in an image plane of the register tweezers minimizes the 423~nm leakage light incident on any of the MZ, IZ, SZ and LZ regions. 


Atoms trapped in the 423~nm register tweezers have an average radial (axial) trap frequency of 56~kHz (12.5~kHz) during the initial loading, rearrangement, and 1Q gates. The atoms in IZ, MZ and SZ have an average radial (axial) trap frequency of 70~kHz (15.5~kHz). The distance between atom pairs is 3.3~$\mu$m~\cite{norcia2024iterative}, and the bottom row of MZ and the top row of register is separated by 33~$\mu$m, leaving 9 empty rows in between to facilitate row-wise atom movements and reduced 423~nm leakage light into IZ and MZ.

Arbitrary connectivity and selection of atoms for MCM is achieved by relocating selected atoms between different zones. Arbitrary 2Q gate connectivity is established by moving atoms between the register and the IZ, while MCM selection is achieved by moving selected atoms from the register to the MZ. 
Moves are performed in parallel between the register and other zones, up to specific constraints (see Ref.~\cite{2024AtomQEC}). All movements within the LZ, MZ, and SZ in this work are executed by a single tweezer. These movements necessitate identifying a target and a source from different zones in real time, making these \textit{conditional movements}. 
During any movement, atoms are transferred between static and rearrangement tweezers by ramping up the depth of the rearrangement tweezers over 0.4 ms.

Our universal gateset is demonstrated in a static architecture in Ref.~\cite{muniz2024gates} and in a zoned architecture in Ref.~\cite{2024AtomQEC}. 
We remark on three aspects of the current architecture: (1) 1Q gates are exclusively performed in the register, with row-wise parallelization.  Each qubit has 1Q gates applied at a single location within the register. (2) CZ gates are performed in parallel on up to eight pairs of arbitrary qubits, and (3) leakage into the $^{3}{\rm P}_0$ and Rydberg states is converted to detectable atom loss after each 2Q gate with probabilities 99\% and 80\% respectively, because the metastable states are not trapped in the 423~nm tweezers, and a fraction of Rydberg population decays to these states  \cite{2024AtomQEC}. Because our imaging independently distinguishes the qubit state from atom loss \cite{norcia2023midcircuit}, this allows for ``delayed-erasure" decoding techniques to be applied \cite{2024AtomQEC,chow2024leakage,perrin2025quantumerrorcorrectionresilient}.  


\begin{table}[]
\caption{Errors and loss during regular and MCM imaging.}
\begin{tabular}{|c|cc|}
\hline 
\multicolumn{1}{|l|}{} & \multicolumn{1}{c|}{Regular imaging}  & MCM imaging \\ \hline \hline
Atom loss / bright image      & \multicolumn{1}{c|}{\textless{}0.0006} & 0.005(2)    \\ \hline
Distinguishability error          & \multicolumn{1}{c|}{0.0003} & 0.003       \\ \hline
$^3$P$_2$ loss / bright image       & \multicolumn{1}{c|}{0.0004(6)}        & 0.005(2)    \\ \hline
$1 \rightarrow 0$  & \multicolumn{2}{c|}{0.003(16)}                      \\ \hline
$0 \rightarrow 1$  & \multicolumn{2}{c|}{0.0006(8)}                      \\ \hline
\end{tabular}
\label{table:spam}
\end{table}

\section{MCM sequence}\label{mcm_sequence}

Determining both the state and presence of a qubit in the MZ requires two consecutive MCM images. The initial image identifies whether the atom is in the $\ket{1}$ state. To distinguish atoms in the $\ket{0}$ state from those that are lost, we collect a subsequent image after an optical pumping pulse that transfers all atoms to $\ket{1}$. (we can also image the $\ket{0}$ state directly, but performing two images of  $\ket{0}$ allows us to cancel light-shifts using spin-echo techniques, described below)

\begin{figure}
    \includegraphics[width=1.0\columnwidth]{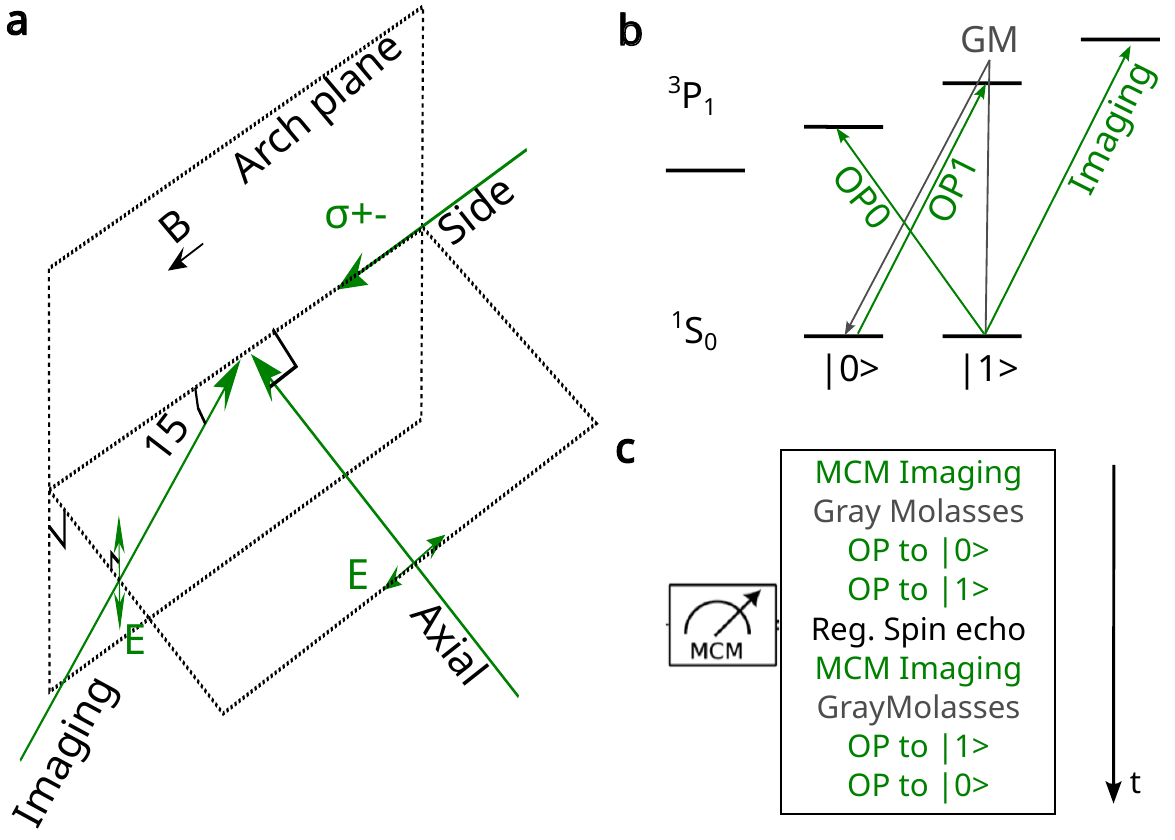}
    \caption{\textbf{a.} MCM beam geometry. ``Arch'' plane is the plane where the zoned architecture is formed. The ``Imaging" beam addresses all zones. ``Side" and ``Axial" beams locally address IZ, SZ, and MZ. There are independent global ``Side" and ``Axial" beams for the regular (non MCM) state preparation and measurement in the system (not shown). \textbf{b.} MCM beam $^3$P$_1$ sublevel addressing scheme. Optical pumping to state $\ket{0}$ and $\ket{1}$ is provided by the ``Side" beam. Gray-molasses (GM) $\sigma^+$ and $\pi$ are provided by the ``Side" and ``Axial", respectively. \textbf{c.} MCM block composition. The MCM sequence is symmetrized about a spin echo pulse in the register, though with the optical pumping order flipped.
    }
    \label{fig:mcm_seq_appendix}
\end{figure}

The detailed beam geometry, level addressing, and sequence composition of the MCM block is shown in Fig.~\ref{fig:mcm_seq_appendix}. Imaging is performed in a cavity-enhanced 784~nm 3D lattice by collecting the photons scattered from the cycling transition from state $\ket{1}$ to $^3$P$_1 \ket{F=3/2, m_F=3/2}$ with a high numerical-aperture microscope objective \cite{norcia2024iterative}. The MCM image is 7~ms long, and the applied 556~nm light has a saturation parameter of approximately 1 and detuning of roughly 100~kHz. Due to the limited optical access, the imaging beam is inserted through one of the in-vacuum cavity mirrors with its k-vector forming a 15~deg angle to the 500 G magnetic field, and illuminates the entire atom array. The beam is linearly polarized orthogonal to the magnetic field, having both $\sigma^+$ and $\sigma^-$ components. Due to the Zeeman splitting of the $^3$P$_1$ sublevels, state-selective imaging of $\ket{1}$ is achieved using only the $\sigma^+$ component. Different frequency tones are generated by a fiber phase electro-optic modulator to reach each transition~\cite{norcia2023midcircuit}. 

Identification of the atomic state and presence is similar to cavity-based imaging presented in our previous works \cite{norcia2024iterative, 2024AtomQEC}.  We identify the state and presence of atoms by comparing the number of photons collected in an image to a threshold.  For MCM, care must be taken to avoid scattering of 423~nm leakage light into the MZ (see Fig. \ref{fig:hiding_shift}), which can degrade imaging performance.  Measurement error information is shown in the Table~\ref{table:spam}. The MCM imaging incurs atom loss of 0.005(2) per image; higher than our non-MCM imaging due to optical pumping into $^3$P$_2$ state by 423~nm leakage light (see~\ref{3_level_system}). A 1388~nm $^3$P$_0$ repump laser is used for both regular and MCM imaging, and it improves the 784~nm cavity imaging loss by 0.0021(7)(0.0014(5)) for regular (MCM) imaging. For MCM images, we compromise state distinguishability slightly by reducing the number of imaging photons scattered in order to reduce atom loss. The spin flip probability from $\ket{0}$ to $\ket{1}$ is much lower than the spin flip probability from  $\ket{1}$ to $\ket{0}$ because the $\ket{0}$ state is not excited to $^3$P$_1$. A background loss of about 0.0002 per image is incurred by our 30~s vacuum lifetime.

Following the MCM image, 4~ms of gray molasses cooling is performed~\cite{jenkins2022yb,2024AtomQEC,li2025parallelized} by addressing the $^3$P$_1$ $\ket{3/2,1/2}$ state with two beams in a Raman configuration ($\sigma^+$ along the magnetic field and $\pi$ through the objective) with a single-photon detuning of $+1.5$~MHz. The relative detuning of the two Raman beams is adjusted to be nearly equal to the qubit frequency splitting to give optimal cooling. Additionally, a 300$\mu$s optical pumping pulse prepares atoms in $\ket{0}$ ($\ket{1}$) by addressing the $^3$P$_1$ $\ket{3/2,-1/2}$ ($\ket{3/2,1/2}$) states with a local $\sigma$-polarized beam (``Side" beam in Fig.~\ref{fig:mcm_seq_appendix}). All beams used for cooling and optical pumping are localized to IZ, MZ, and SZ, though some leakage light into the register region is present. During the MCM cooling and optical pumping, atoms are simultaneously held in the optical cavity lattice and 423~nm tweezers, mitigating the effects of leakage light. 

Given that our imaging protocol utilizes global 556~nm beams, qubits held within the register experience position-dependent inhomogeneous phase shift as in Ref.~\cite{norcia2023midcircuit}. To counteract this effect, we implement 1Q BB1 composite echo pulse on the qubits in the register \cite{wimperis1994BB1}. After each MCM image, cooling and optical pumping are applied in order to ensure that any induced shifts cancel out in the echo sequence. After the first image, optical pumping places atoms first into $\ket{0}$ and then into $\ket{1}$. After the second image, the order is reversed, preparing atoms in the SZ in the $\ket{0}$ state. Crucially, atoms in $\ket{0}$ are not promoted to the Rydberg state during subsequent 2Q gate operations, so atoms in the SZ do not interfere with gates in the nearby IZ.

After all atoms in both the MZ and the SZ are cooled and prepared in the $\ket{0}$ state, a software service identifies the locations of the remaining atoms in the SZ and lost atoms in the MZ. The service then streams real-time instructions to individually move the necessary atoms, correcting for detected loss events. The latency associated with this software service depends on the location and number of atoms to move, but for our typical circuits is consistently below 10~ms. We routinely achieve SZ-to-MZ conditional moves efficiencies exceeding 99.6\%, measured  through repeated cycles of populating one zone from the other using our standard MCM cycles. 

\section{Light-shifts and loss channels from register tweezer light}\label{3_level_system}

In order to shield atoms in the register from light used in the MCM sequence (either due to the global imaging beam or leakage from the local cooling beams), we operate the register tweezers near the $^3$P$_1$ - $8s^3$S$_1$ transition, which creates large shifts on the $^3$P$_1$ manifold. At our chosen detuning, the shifts on the $^3$P$_1$ $\ket{3/2,+3/2}$ state (imaging) are approximately $2\times$ larger than the shifts on the $\ket{3/2,\pm 1/2}$ states (cooling and optical pumping). We increase the register power by 33\% for the cooling step to balance the required shifts against the negative effects of light leakage. We also choose the optical power to avoid any accidental resonance product of the different frequency sidebands present after light is modulated \cite{norcia2023midcircuit}.

We operate the register traps with a detuning of $\Delta_{3S1} \simeq 2\pi \times -7.6$~GHz from the $^3$P$_1$ $\ket{3/2,+3/2}$ to $8s^3$S$_1$ $\ket{3/2,+3/2}$ transition with a Rabi frequency of $\Omega_{423} = 2\pi \times 2.4 $ GHz during imaging. Although this provides an AC Stark shift of nearly 100~MHz on the imaged state, the combination of both the imaging light and register trap laser can cause the atoms to populate $8s^3$S$_1$ via a detuned two-photon process, which subsequently populates the long-lived $^3$P states after spontaneous decay. This decay primarily populates the $^3$P$_2$ state, which is difficult to repump due to its large number of sublevels that are widely dispersed in energy by our large magnetic field. These atoms are eventually lost as our 459~nm traps do not trap the $^3$P$_2$ state. However, we are able to repump some small population from $^3$P$_0$, estimated to be less than 10\% of the all leakage events caused by 423~nm.

\begin{figure}
    \centering
    \includegraphics[width=\linewidth]{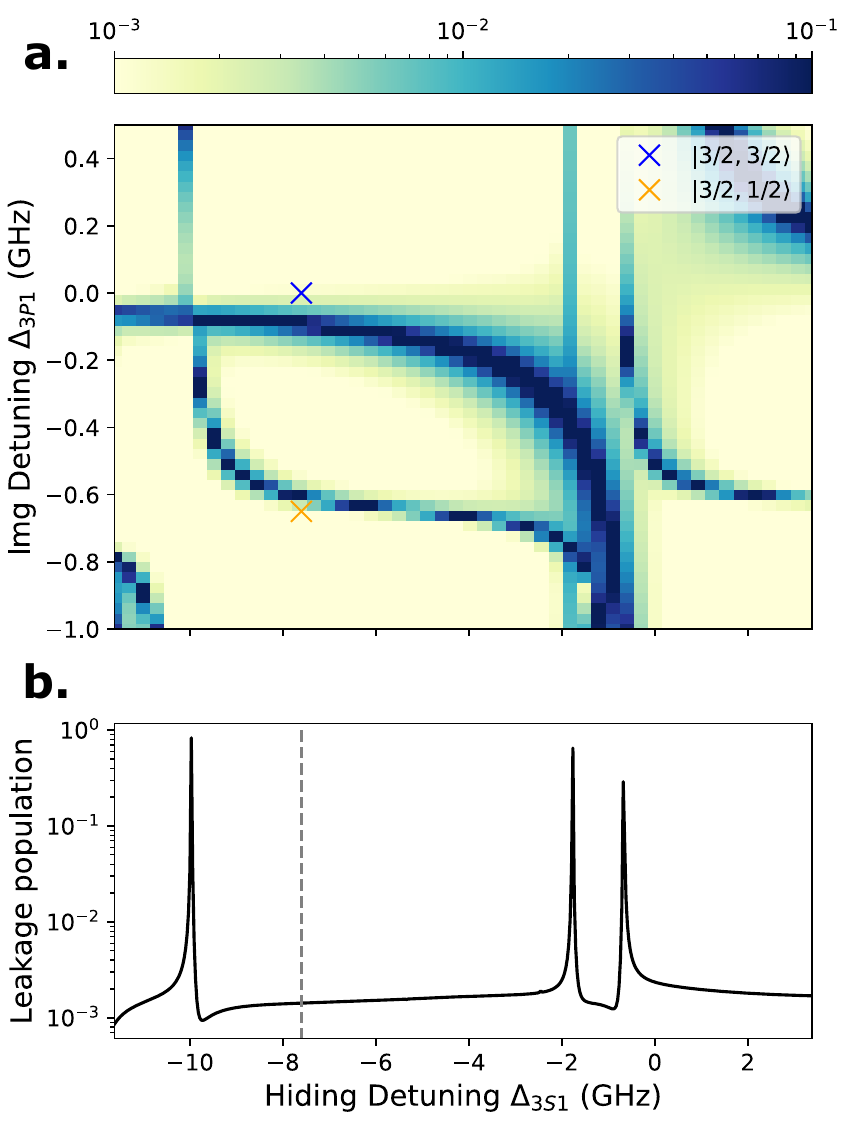}
    \caption{Loss of register atoms during imaging.  a) Calculated per-image loss for our typical MCM imaging parameters (see text) for atoms in the register, versus the register laser frequency (horizontal axis) and imaging laser frequency (vertical axis). Crosses represent our operating conditions for imaging (blue) and cooling (orange). Note that the scattering rate depicted here at the location of the orange cross is not relevant, as the polarization and intensity used during cooling are different from imaging. b) A slice through the same data near our imaging detuning $\Delta_{3P1} = 0$. The dashed line represents our detuning $\Delta_{3S1}$, indicating an expected loss near $10^{-3}$ for register atoms during an image. The resonances correspond to the different hyperfine levels of $8s^3$S$_1$.}
    \label{fig:hiding_shift}
\end{figure}

To understand the scattering of the atoms during the hiding process, we construct a 16-level model including all hyperfine levels of $^1$S$_0$, $^3$P$_1$, and $8s^3$S$_1$, as well as additional leakage states to represent the long-lived $^3$P states where the atoms are lost. We then numerically solve the optical Bloch equations for both the imaging and cooling processes assuming decay from the $^3$P$_1$ state the ground state as well as decay from the $^3$S$_1$ state to both $^3$P$_1$ and the leakage state to calculate the population in the leakage state after these processes. A summary of the calculated populations during these different processes can be found in Table \ref{tab:hiding_loss_channels}. As shown in Fig.~\ref{fig:hiding_shift}(a), the presence of the near-resonant 423~nm light shifts the resonances where the loss is enhanced by approximately $-100$~MHz (+50 MHz) for the $\ket{3/2,3/2}$ ($\ket{3/2,1/2}$), matching the measured Stark shift at this Rabi frequency. The crosses represent the operating points for the imaging and cooling respectively, matching the unshifted resonances for those states. In Fig.~\ref{fig:hiding_shift}(b) we show a slice of the diagram at $\Delta_{3P1} = 0$, indicating that at our chosen detuning we will lose approximately $0.0015$ of the register atoms during the imaging process. Given that each MCM cycle consists of two images, we expect this to contribute $0.003$ to the total MCM loss for atoms in the register. Additional tones on the imaging light due to the use of an electro-optic modulator (EOM) to generate the imaging tone, and the presence of RF spurs on the EOM drive may contribute to additional loss.  

Additional loss of atoms in the register is contributed by the cooling process. In particular, we find that loss in the register mostly depends on the overlap of the local $\sigma^+$ polarized beam near the $\ket{3/2, 1/2}$ state that is used during cooling in the MZ and SZ. Simulations indicate that this polarization provides multiple loss channels through different intermediate states, including a transition to the $8s^3$S$_1$ state with less than 2 GHz of detuning. While this beam is nominally local to the measurement zone, we measure approximately $10\%$ of the intensity incident on the register, leading to approximately a loss of $0.003$ per cooling cycle ($0.006$ loss per MCM cycle). 
This loss channel in conjunction with the imaging loss described above accounts for the observed loss of register atoms described in the main text. Future work will improve the extinction ratio of the local beams to significantly reduce this effect.

\begin{table}
    \begin{tabular}{|c|c|c|}
        \hline
         & Register & MZ \\
         \hline\hline
        Imaging & 0.003 & 0.005 \\
        Cooling & 0.006 & 0.00005\\
        \hline
        \textbf{Theory Total} & 0.009& 0.005  \\
        \hline
        \textbf{Exp Meas} & 0.0106(7) & 0.005(2)\\
        \hline
    \end{tabular}
    \caption{Summary table describing the sources and proportion of lost atoms during the MCM cycle. The calculated loss during the imaging and cooling processes accounts for the vast majority of observed loss in both the register and the MZ as measured in the main text.}
    \label{tab:hiding_loss_channels}
\end{table}

Even small amounts of 423~nm light leaking into the MZ during MCM operations can lead to population of the higher lying $^3$S$_1$ state that quickly decays to the longer lived $^3$P manifold.
This is the main factor that leads to loss for atoms in the MZ during MCM cycles. This light leakage could be due to scattering in some of the many glass surfaces used in the optical path, as even a hard stop in an earlier image plane does not completely extinguish it. 

We calculate that a extinction ratio of $1:10^{-5}$ (corresponding to an experimental bound on this leakage) for 423~nm light on the MZ can lead to an atom loss of 0.005 into the $^3$P manifold after a $t_{img}= $7~ms imaging pulse. 
This level of leakage light contributes a negligible Stark shift of less than $1\%$ of the imaging Rabi frequency so does not appreciably modify the $^3$P$_1$ state population during imaging. However, this light off-resonantly drives atoms in the $^3$P$_1$ state to the the leakage manifold via the $^3$S$_1$ state with a population proportional to $10^{-5}(\Omega_{423}/\Delta_{3S1})^2 \times \Gamma_{3S1}t_{img}$, leading to a loss rate of approximately $0.0025$ per image ($0.005$ per MCM cycle). This can be reduced by increasing the detuning $\Delta$ or lower the overall power of the tweezers to reduce $\Omega_{423,MZ}$, but at fixed contrast, this increases error rates on register atoms during MCM.

\section{Atom replenishment sequence}\label{reloading}
The operation of our MOT and transport lattice are described in previous work \cite{norcia2024iterative}, with the LZ referred to in this work corresponding to the reservoir of reference \cite{norcia2024iterative}.  
As before, to minimize 532~nm light incident on the data qubits stored in the register, the LZ is loaded with atoms from the transport lattice at a distance of 170~um, and once loaded the whole array is moved near the other arrays using a galvo mirror in a conjugate plane. 
The LZ used in this work has 75 sites but no highways, such that the atom packing is denser (3.3$\mu$m intersite). 

We bound the decoherence contributed by operation of the MOT by measuring the decay in contrast in a Ramsey echo experiment with a MOT and without the MOT lasers on during the total hold time. 
If a degradation of contrast occurs due to the formation of a MOT, we would expect it to have an exponential form versus time, as the most likely culprits would be associated with scattered light (the magnetic field environment does not change whether we form a MOT or not).
From an exponential fit to contrast decay out to 2 seconds hold time, we observe a contrast decay rate of 0.029(3)/s and 0.035(4)/s without the MOT and with the MOT, respectively, corresponding to a differential contrast decay of 0.006(5)/s. This rate is negligible compared to other sources of error while running circuits, so the MOT can be loaded in parallel with other operations without meaningfully impacting performance. In contrast to reference \cite{norcia2023midcircuit}, we use a core-shell MOT configuration \cite{Lee2015coreshell}, which drastically reduces 399~nm scattering from cooled atoms while maintaining a fast loading rate. 

Given the low impact of the MOT during circuit execution (see below), we operate the MOT continuously in parallel with the circuits. The MOT is turned off only during the transport step while loading the atoms in the 532~nm conveyor belt lattice. The lattice takes 120~ms to transport the atoms 300~mm vertically from the MOT chamber to the science chamber. During the subsequent handoff, the lattice is ramped down over 10~ms while the loading tweezers are ramped up simultaneously. Finally, the galvo movement takes 10~ms approximately, leaving the array $\sim 6.6$~{\textmu}m from the SZ.

Before rearranging atoms from the LZ to the SZ, we perform a parity projection step via light-assisted collisions (LACs) that creates an array with single atoms and 50\% loading fraction on average \cite{schlosser2001sub}. The LAC beams are local to the LZ, SZ, MZ and IZ, and address the $^3P_1 \ket{3/2, \pm1/2}$ states.  
After single atoms are prepared, we use an MCM cycle to image and cool atoms in the LZ and SZ. Using these images, the software service calculates the rearrangement movements from the LZ to refill the SZ. The duration of these movements is non-deterministic as it depends on the number of sites to refill and its positions, but on average a full refilling of the SZ typically takes about 120 ms, including the software latency from the move compilation.

\begin{figure}
    \centering
        \includegraphics[width=1.0\columnwidth]{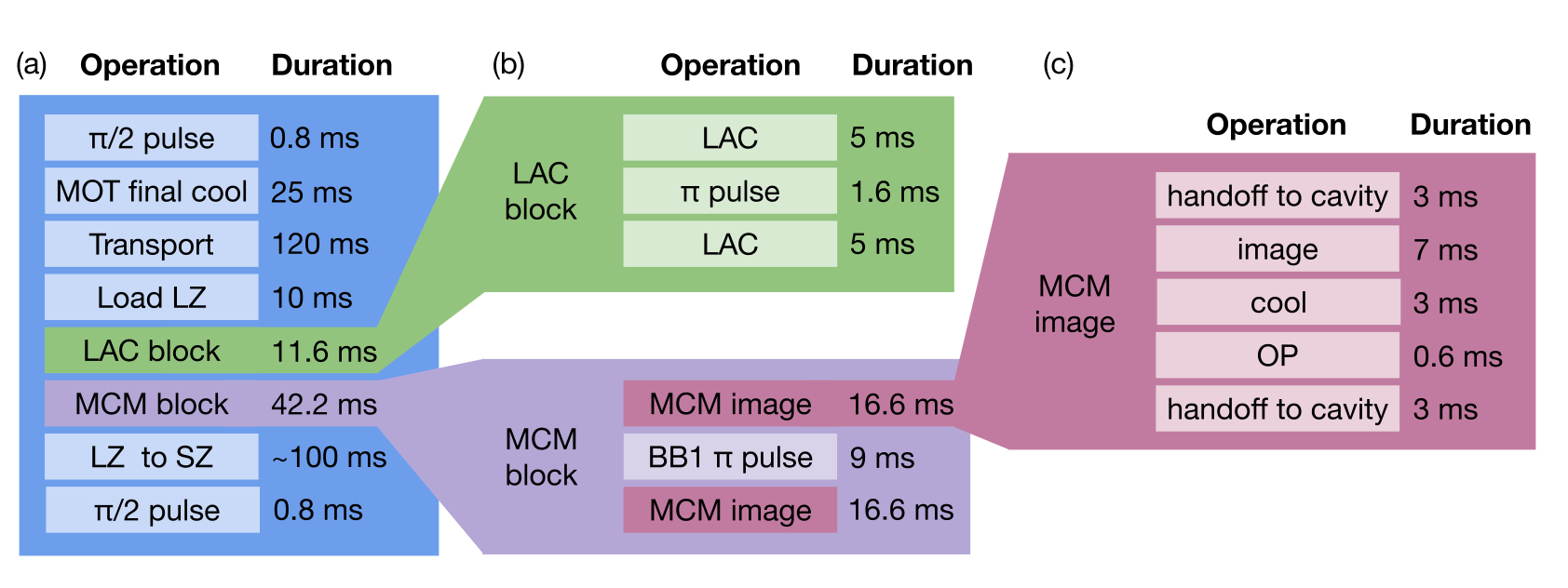}
    \caption{Timing for mid-circuit reloading sequence.  See text in Appendix \ref{reloading} for details.  The $\pi/2$ pulses surrounding the reloading operations are included for characterization of decoherence caused by the reloading sequence.  1Q gates indicated are applied to atoms in the register.    }
    \label{fig:reloading_timing}
\end{figure}





\section{Repetition code with ancilla reuse} \label{rep_code_extra}

In a standard circuit-based repetition code memory experiment, the data qubits are not measured until the end of the circuit. This construction prevents the detection of qubit loss which is detected only when qubits are measured. Additionally, our atoms experience heating that scales with gate depth and movement count, so it is desirable to regularly image and reset the atoms to cool them.

To ensure that every atom is measured frequently, we make several changes from a standard repetition code. First, we add an additional parity check, turning the repetition code into a 1D toric code, or ``ring code.'' Second, we compile SWAP gates into the syndrome extraction cycle, converting between syndrome qubits and data qubits every cycle \cite{2024AtomQEC,chow2024leakage,perrin2025quantumerrorcorrectionresilient}. After compilation to CZ gates, this SWAP operation adds no extra 2Q gates compared to a standard syndrome extraction circuit and requires only a possible Pauli correction that can be implemented in software. Furthermore, this change ensures that no atom remains active for more than 4 2Q gates, in turn allowing us to detect lost atoms and cool those that remain. We implement SWAP operations so that the information stored in qubit $i$ at the beginning of one cycle is transferred to qubit $i+1 \mod 2n$ at the end of the cycle. Thus, the quantum information gradually cycles around the ring; we refer to this circuit as a walking repetition code. 

\begin{figure*}
    \centering
        \includegraphics[width=2.0\columnwidth]{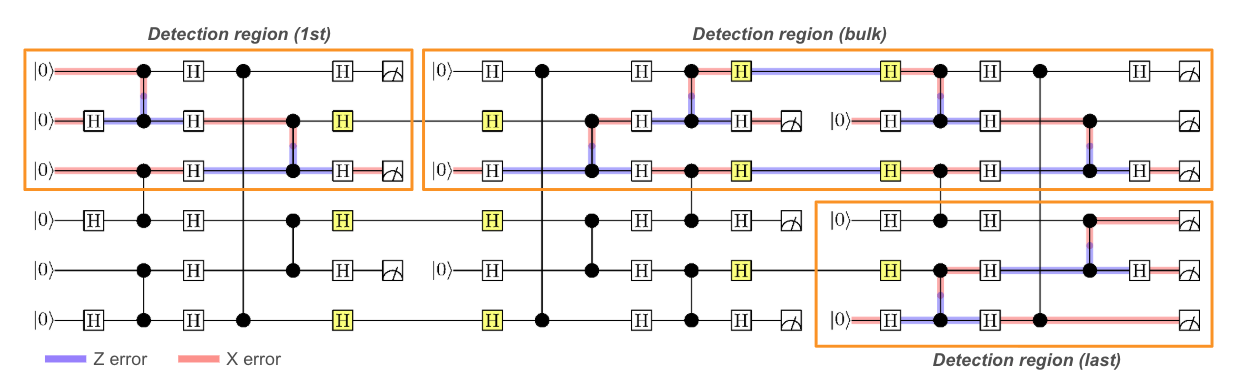}
    \caption{ Example detection regions for a distance-3 phase-sensitive repetition code. We show sensitivity for both $Z$-type (blue) and $X$-type (red) Pauli errors. Any single Pauli error of appropriate type changes the parity of the corresponding detector. The phase-insensitive version of the circuit excludes the Hadamard gates (highlighted) on idle qubits surrounding the MCMs.}
    \label{fig:rep_code_appendix}
\end{figure*}

Following Refs. \cite{fowler2014scalable,googleAI_repcode_2021}, we define `detectors` as sets of measurements with determined parity in the absence of error. For our walking repetition code, a parity check from measuring qubit $j$ at the end of one cycle will correspond to a parity check measured on qubit $j + 1 \mod 2n$ at the end of the subsequent cycle. The set of two such measurements (in the bulk) defines a detector. In the first cycle and for the final measurement, we define similar detectors with known parity by comparing to known initial or final states. Defining detectors in this manner creates localized regions of error sensitivity, as shown in Fig. \ref{fig:rep_code_appendix}. To compute detector values in the presence of atom loss, we populate bits corresponding to lost atoms with random values. 

Since the repetition code only protects classical information, the detection regions are not sensitive to all quantum channels. Since we anticipate $Z$-biased noise on idle qubits during MCM, we construct two different versions of the repetition code circuits, a phase-insensitive version that does not detect phase errors during MCM and a phase-sensitive version that detects phase errors during MCM. As expected, we see increased detection frequency for the phase-sensitive version as discussed in the main text. 

In order to decode these results, we used the sparse-blossom variant of the minimum-weight perfect matching algorithm, implemented by PyMatching \cite{Higgott2025sparseblossom}. The initial matching graph was generated using Stim based on a predetermined error model \cite{Gidney2021stimfaststabilizer}. The noise parameters used in this model were initially based on fidelity estimates and gate characterizations, and were then fine-tuned over earlier runs of repetition code experiments. For the data shown in this paper, we used the final set of fine-tuned parameters. 

To account for atom loss, the matching graph was edited on a per-shot basis, using techniques similar to those described in Ref. \cite{baranes2025leveraging}. For each detected atom loss, we modify the edge weights in the matching graph in two ways. First, we update timelike edges (corresponding to readout errors) to have zero-weight when connected to detectors affected by atom loss. Second, we modify edge weights corresponding to Pauli probabilities on partner qubits sharing CZ gates with a potentially lost atom. We call these errors ``loss-correlated gate errors''. A comparison between various decoders will be discussed in future work.

In Table \ref{tab:rep_data}, we provide the data used to generate Fig. \ref{fig:rep_code}(c).

\begin{table}[]
    \centering
    \caption{Decoding failure data for repetition code experiment runs.}
    \begin{tabular}{|c|c|c|}
    \hline
     Distance & Logical Failures  & Total Shots\\ \hline \hline
      \multicolumn{3}{|c|}{Phase-sensitive} \\ \hline
      3 & 30 & 11700\\ \hline
          5 & 11 & 11250\\ \hline
          7 & 6 & 11250\\ \hline
        \multicolumn{3}{|c|}{Phase-insensitive} \\ \hline
      3 & 25 & 17820\\ \hline
          5 & 3 & 17640\\ \hline
          7 & 4 & 17460\\ \hline
          9 & 3 & 16920\\ \hline
    \end{tabular}
    \label{tab:rep_data}
\end{table}

\section{Encoded Bell pair distillation}\label{422_prep}

\begin{figure}
    \centering
        \includegraphics[width=0.3\columnwidth]{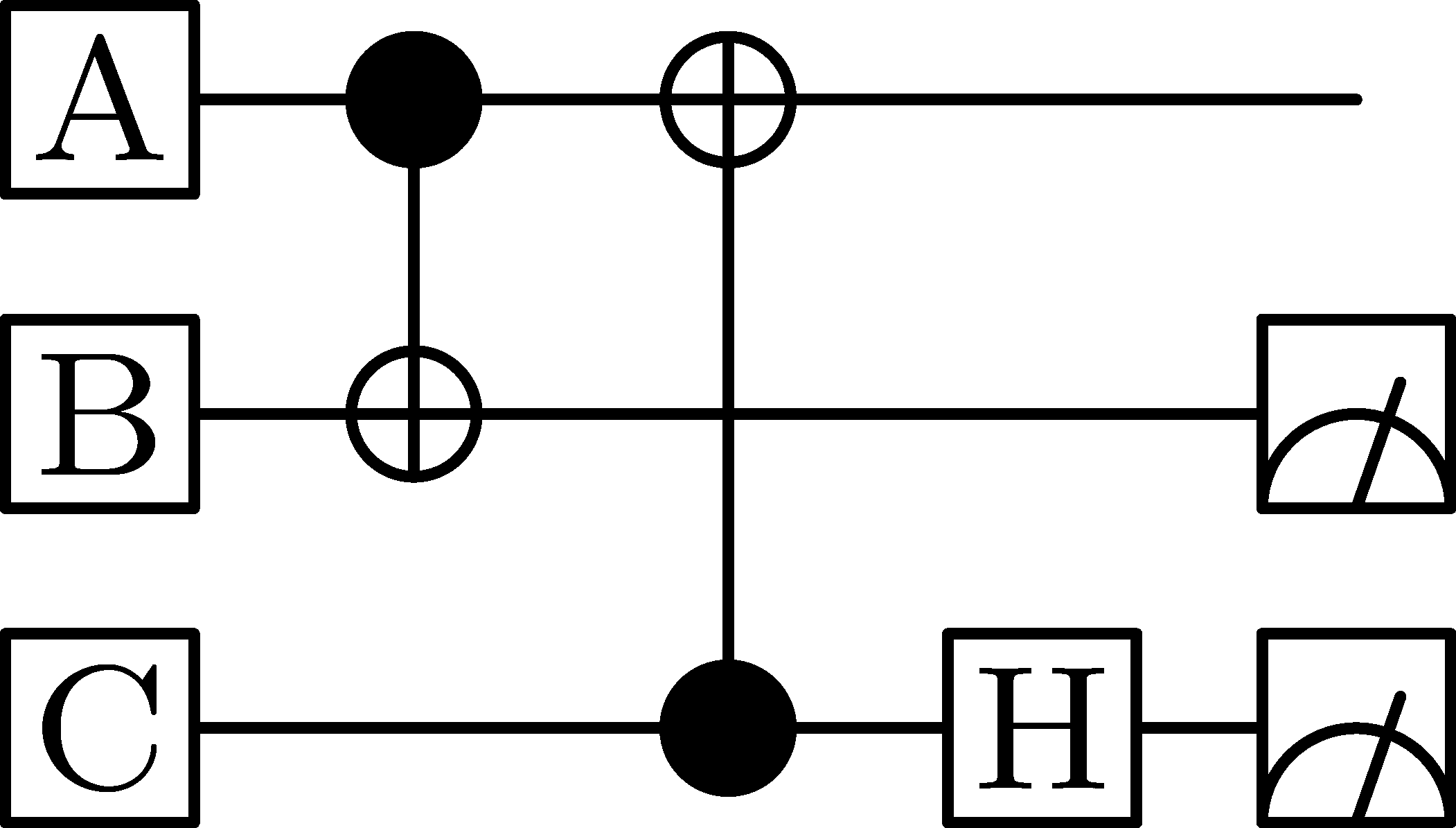}
    \caption{A schematic of the [[2,1,2]] distillation protocol. Blocks A, B, C represent Bell pairs, either encoded or unencoded. The CNOT gates are applied in a transversal manner between the blocks. The procedure is complete if both blocks B and C are measured in even parity states.}
    \label{fig:distillation}
\end{figure}

The [[2,1,2]] distillation protocol \cite{deutsch1996} is illustrated in Fig.~\ref{fig:distillation}. To detect bit-flip errors during preparation of a Bell pair, a second Bell pair is also prepared. A transversal CNOT gate is applied between the two Bell pairs, propagating bit-flip errors from the control block to the target block. Measuring the target block in the incorrect state (odd parity) indicates an error during the preparation of one of the Bell pairs. This information is insufficient to determine where the error occurred, so if the measurement indicates an error, both blocks are discarded. This procedure suppresses physical errors at first order. An analogous procedure can be applied to detect phase errors. 

When operating on physical Bell pairs, this procedure is not tolerant to gate faults. To make the procedure fault-tolerant requires encoding the procedure into logical blocks. In this work, we prepare logical Bell pairs using [[4,2,2]] codes \cite{Vaidman1996code,Grassi1997code,linke2017FTdetection}. This code is a stabilizer code \cite{Gottesman1998FTtheory} that can detect any 1Q Pauli error and correct for any single-qubit loss event. The [[4,2,2]] code stores two logical qubits in four physical qubits and admits a transversal CNOT operation that acts as logical CNOT gates between corresponding logical qubits across two [[4,2,2]] code blocks.  

To implement an encoded distillation scheme, we prepare three [[4,2,2]] code blocks in logical Bell states. The preparation of physical states $\ket{\Phi^+}\otimes\ket{\Phi^+}$ fault-tolerantly prepares an undistilled logical Bell pair $\ket{\bar{\Phi}^+}$ in the [[4,2,2]] code, as demonstrated previously in Ref.~\cite{2024AtomQEC}. We treat one block as the ``data block'' while the other two blocks are used to detect bit-flip and phase errors. The full protocol prepares logical Bell states in a manner that suppresses any single Pauli error. After the ancillary blocks are measured using MCM, the same software service that deals with conditional movements between zones decodes the ancilla measurement. Following this check, we rotate the data qubits in order to measure the logical qubits in the $XX$, $YY$, or $ZZ$ basis. The loop exit is triggered by a successful check, after which the data qubits are measured in the register. If the state preparation is not successful, all physical qubits are reset via another MCM block and the procedure starts over. Note that the final measurement of this procedure reintroduces errors that are detectable but not suppressed by the distillation procedure.

In addition to the procedure described above, we performed a variant in which anti-ferromagnetic states were created. This corresponds to the preparation of entangled states $|\Psi^-\rangle$. Two copies of this physical state produce an encoded state $|\bar{\Psi}^-\rangle$. These states are insensitive to global coherent phase noise, even when physical qubits are rotated using Hadamard gates. We do not observe any significant difference when performing these experiments, so we do not distinguish among them in the analysis shown here.

Here we include the raw experimental data used to calculate the fidelities for encoded and unencoded distillation experiments reported in the main text in Table \ref{tab:belldata}.

\begin{table}[]
    \centering
    \caption{Measurement results for unencoded and encoded Bell pair distillation. Here, a failure corresponds to an incorrect parity measurement for the target Bell state.}
    \begin{tabular}{|c|c|c|c|}
    \hline
     & $XX$ basis & $YY$ basis  & $ZZ$ basis\\ \hline \hline
     \multicolumn{4}{|c|}{Unencoded} \\ \hline
     Successes & 2443 & 2127 & 4828\\ \hline
     Failures & 7 & 61 & 79\\ \hline
     \multicolumn{4}{|c|}{Encoded} \\ \hline
     Successes & 1440 & 1366 & 1452\\ \hline
     Failures & 3 & 3 & 4\\ \hline
    \end{tabular}
    \label{tab:belldata}
\end{table}


\bibliography{bib}
\end{document}